\documentclass[aps,onecolumn,12pt]{revtex4}
\usepackage{amsmath}

\begin{document}
\title{Is the Cosmological Constant Problem Properly Posed?}
\author{Philip D. Mannheim}
\affiliation{Department of Physics, University of Connecticut, Storrs, CT 06269, USA.
email: philip.mannheim@uconn.edu}
\date{March 27, 2017}
\begin{abstract}
In applications of Einstein gravity one replaces the quantum-mechanical energy-momentum tensor of sources such as the degenerate electrons in a white dwarf or the black-body photons in the microwave background by c-number matrix elements. And not only that, one ignores the zero-point fluctuations in these sources by only retaining the normal-ordered parts of those matrix elements. 
There is no apparent justification for this procedure, and we show that it is precisely this procedure that leads to the cosmological constant problem. We suggest that solving the problem requires that gravity be treated just as quantum-mechanically as the sources to which it couples, and show that one can then solve the cosmological constant problem if one replaces Einstein gravity by the fully quantum-mechanically consistent conformal gravity theory.

\end{abstract}
\maketitle

\vskip3.0truein

\noindent
Essay written for the Gravity Research Foundation 2017 Awards for Essays on Gravitation.

\newpage
\section{Origin of the Cosmological Constant Problem}

Despite their familiarity, it is the very way in which the standard Einstein gravitational equations with matter source ${\rm M}$, viz.
\begin{eqnarray}
\frac{1}{8\pi G}\left(R^{\mu\nu} -\frac{1}{2}g^{\mu\nu}R^{\alpha}_{\phantom{\alpha}\alpha}\right)=-T^{\mu\nu}_{\rm M}, 
\label{M166}
\end{eqnarray}
are used in astrophysics and cosmology that actually creates the cosmological constant problem. To establish this we start by noting that since the two sides of the equation are to be equal to each other, the two sides must either both be quantum-mechanical or must both be classical.  However, since the gravity side is not well-defined quantum-mechanically, one takes the gravity side to be classical. Now at the time the Einstein equations were first introduced the energy-momentum tensor side was taken to be classical too. However, with electron degeneracy being able to stabilize a white dwarf star up to an expressly quantum-mechanically-dependent Chandrasekhar mass of order $(\hbar c/G)^{3/2}/m_p^2$, and with the cosmic microwave background being a black-body with energy density equal to $\pi^2k_{\rm B}^4T^4/15\hbar^3c^3$, it became clear not just that quantum mechanics is relevant on large distance scales, but that gravity is aware of this, and that the quantum-mechanical nature of its macroscopic sources is relevant to gravitational astrophysics and cosmology.

To try to get round the fact that the gravity side of the Einstein equations is classical (${\rm CL}$) while the matter side is quantum-mechanical,  one replaces the quantum-mechanical $T_{\rm M}^{\mu\nu}$ by its c-number matrix elements in appropriate states $|\psi\rangle$, and thus replaces (\ref{M166}) by 
\begin{eqnarray}
\frac{1}{8\pi G}\left(R^{\mu\nu} -\frac{1}{2}g^{\mu\nu}R^{\alpha}_{\phantom{\alpha}\alpha}\right)_{\rm CL}=-\langle \psi|T^{\mu\nu}_{\rm M}|\psi \rangle.
\label{M167}
\end{eqnarray}
Now since the matter term in  (\ref{M167}) consists of products of quantum fields at the same spacetime point, the matter term has an infinite zero-point contribution $(\sim \hbar \int d^3kk^{\mu}k^{\nu}/\omega_k$).  But with the gravity side of (\ref{M167}) being finite, it cannot be equal to something that is infinite. Thus one must find a mechanism to cancel infinities on the matter side, and must find one that does so via the matter side alone. However instead, in the literature one commonly ignores the fact that the hallmark of Einstein gravity is that gravity is to couple to all forms of energy density rather than only to energy density differences, and subtracts off (i.e. normal orders away) zero-point infinities by hand, and replaces (\ref{M167}) by the finite (${\rm FIN}$) 
\begin{eqnarray}
\frac{1}{8\pi G}\left(R^{\mu\nu} -\frac{1}{2}g^{\mu\nu}R^{\alpha}_{\phantom{\alpha}\alpha}\right)_{\rm CL}=-\left(\langle \psi|T^{\mu\nu}_{\rm M}|\psi \rangle\right)_{\rm FIN}.
\label{M168}
\end{eqnarray}
Thus in treating the contribution of the electron Fermi sea to white dwarf stars or the contribution of black-body photons to cosmic evolution, one uses an energy operator of the generic form $H=\sum(a^{\dagger}(\bar{k})a(\bar{k})+1/2)\hbar\omega_k$, and then by hand discards the $H=\sum \hbar \omega_k/2 $ term. And then, after all this is done, the finite parts of $\langle \psi |T^{\mu\nu}_{\rm M}|\psi \rangle$ and the vacuum $\langle \Omega|T^{\mu\nu}_{\rm M}|\Omega \rangle$ still have an uncanceled and as yet uncontrolled cosmological constant contribution ($T^{\mu\nu}_{\rm M}\sim \Lambda g^{\mu\nu}$) that needs to be dealt with. Because of their differing structure, the zero-point and cosmological constant terms are distinct, with both problems thus needing to be dealt with.

There would not appear to be any formal derivation of (\ref{M168}) in the literature that starts from a consistent quantum gravity theory \cite{footnote1}, and since it is (\ref{M168}) that is conventionally used in astrophysics and cosmology, it would not appear to yet be on a fully secure footing.  While a derivation of (\ref{M168}) might eventually be forthcoming, in the current gravity literature one starts with (\ref{M168}) as a given,  and then tries to solve the cosmological constant problem associated with the fact that quantum-field-theoretic contributions to the right-hand side of (\ref{M168}) are at least 60 orders of magnitude larger than the cosmology associated with (\ref{M168}) could possibly tolerate \cite{footnote2}. It appears to us that, as currently understood, the standard gravity cosmological constant problem is not properly posed, as it is based on trying to make sense of a starting point for which there would not appear to be any clear justification. Moreover, as written, (\ref{M168}) entails that gravity itself is to play no role in solving the cosmological constant problem as all it can do is respond to whatever energy density the right hand side of (\ref{M168}) provides it with. To give gravity a role it would need to be as quantum-mechanical as the source to which it couples, something that one should anyway want of a physical theory. On making gravity quantum mechanical, below we find that the zero-point problem  and the cosmological constant problem are then tied together and solved together.

\section{On the Nature of Quantum Conformal Gravity}

In quantum electrodynamics (QED) one also has to deal with both classical and quantum-mechanical equations of motion. However, because QED is renormalizable, one can derive classical electrodynamics (CED)  from QED by taking matrix elements of quantum fields in configurations with an indefinite number of photons. One thus has no need to posit an electrodynamic analog of an equation such as (\ref{M167}) since in QED one can actually derive it \cite{footnote3}. In order  to address the cosmological constant problem we would first need to get a justifiable starting point that could then be used for both classical and quantum gravity. Thus, just as in QED, we would need to begin with a renormalizable quantum gravity theory. We are this naturally led to consider the renormalizable conformal gravity theory (see e.g. \cite{Mannheim2006,Mannheim2012,Mannheim2015,Mannheim2017} for some recent reviews).  Moreover, a conformal structure is also natural for the matter sector to which gravity couples,  since the matter sector will also be locally conformal invariant at the level of the action if there are no fundamental mass scales and all mass is generated by spontaneous symmetry breaking.

In conformal gravity the gravitational sector action is taken to be of the form 
\begin{eqnarray}
I_{\rm W}&=&-\alpha_g\int d^4x (-g)^{1/2}C_{\lambda\mu\nu\kappa} C^{\lambda\mu\nu\kappa},
\label{M171}
\end{eqnarray}
where the coupling constant $\alpha_g$ is dimensionless  and $C_{\lambda\mu\nu\kappa}$ is the Weyl conformal tensor. The $I_{\rm W}$ action is the only pure gravity action in four spacetime dimensions that is left invariant under local conformal transformations of the form $g_{\mu\nu}(x)\rightarrow \exp[2\alpha(x)]g_{\mu\nu}(x)$ \cite{footnote4}, and it is because $\alpha_g$ is dimensionless that conformal gravity is renormalizable. 
Now since its equations of motion are fourth-order it had been thought that the theory has states of negative norm. However, on explicitly constructing the quantum Hilbert space it was found \cite{Bender2008a,Bender2008b} that that there were no states of negative norm (and no states of negative energy either) \cite{footnote5}. Conformal gravity is thus offered as a fully consistent quantum theory of gravity, one with no need for the extra dimensions required of string theory.

\section{Solution to the Cosmological Constant Problem}

With conformal gravity being consistent at the quantum level, if we introduce a conformal invariant matter action $I_{\rm M}$, we can take the action of the universe to be the fully conformal invariant $I_{\rm UNIV}=I_{\rm W}+I_{\rm M}$. In the same way that we define the variation of $I_{\rm M}$ with respect to the metric to be $T^{\mu\nu}_{\rm M}$, we can define the variation of $I_{\rm W}$ with respect to the metric to be the gravitational energy-momentum  tensor $T^{\mu\nu}_{\rm GRAV}$,  and can define the variation of $I_{\rm UNIV}$ with respect to the metric to be $T^{\mu\nu}_{\rm UNIV}$. Stationarity of $I_{\rm UNIV}$ with respect to the metric yields $T^{\mu\nu}_{\rm UNIV}=0$, and thus 
\begin{eqnarray}
T^{\mu\nu}_{\rm UNIV}=T^{\mu\nu}_{\rm GRAV}+T^{\mu\nu}_{\rm M}=0.
\label{M174}
\end{eqnarray}
With both $I_{\rm W}$ and $I_{\rm M}$ being renormalizable, the stationarity condition $T^{\mu\nu}_{\rm UNIV}=0$ is not modified by radiative corrections, and thus, in analog to QED, the relation $T^{\mu\nu}_{\rm GRAV}=-T^{\mu\nu}_{\rm M}$ holds both for quantum fields and their c-number matrix elements. Now we had noted above that $T^{\mu\nu}_{\rm M}$ possesses a zero-point term, possessing one even if the matter fields are massless and the vacuum is unbroken. Thus on quantizing the gravitational field, $T^{\mu\nu}_{\rm GRAV}$ must not only possess one too, it must be quantized so that $T^{\mu\nu}_{\rm GRAV}+T^{\mu\nu}_{\rm M}$ is zero-point free \cite{footnote6}. 

Now when particle masses are generated dynamically, a cosmological constant term is induced, and at the same time the matter source zero-point fluctuations readjust as they are now due to vacuum loop diagrams with massive fields. However, since the condition $T^{\mu\nu}_{\rm GRAV}+T^{\mu\nu}_{\rm M}=0$ is an operator identity, it will hold whether the vacuum is a normal one $|\Omega_0\rangle$ or a spontaneously broken one $|\Omega_M\rangle$ in which mass is generated dynamically through the non-vanishing of $\langle \Omega_M|S|\Omega_M\rangle$ where $S$ is an appropriate symmetry breaking field. The wave function renormalization constant of the gravitational field must thus readjust \cite{Mannheim2012,Mannheim2015,Mannheim2017} so that the zero-point and cosmological terms contained in $\langle \Omega_M|T^{\mu\nu}_{\rm M}|\Omega_M\rangle$ are cancelled identically by $\langle \Omega_M|T^{\mu\nu}_{\rm GRAV}|\Omega_M\rangle$. Since all of the infinities in $T^{\mu\nu}_{\rm GRAV}$ and $T^{\mu\nu}_{\rm M}$  are due to the infinite number of modes in the vacuum sector, we can decompose $T^{\mu\nu}_{\rm GRAV}$ and $T^{\mu\nu}_{\rm M}$ into  a divergent contribution from the vacuum  (VAC) and a finite  contribution due to particles that are excited out of the vacuum (PART). Thus on setting $T^{\mu\nu}_{\rm GRAV}=(T^{\mu\nu}_{\rm GRAV})_{\rm VAC}+(T^{\mu\nu}_{\rm GRAV})_{\rm PART}$, $T^{\mu\nu}_{\rm M}=(T^{\mu\nu}_{\rm M})_{\rm VAC}+(T^{\mu\nu}_{\rm M})_{\rm PART}$, (\ref{M174}) will decompose into 
\begin{eqnarray}
(T^{\mu\nu}_{\rm GRAV})_{\rm VAC}+(T^{\mu\nu}_{\rm M})_{\rm VAC}&=&0,
\label{M183}
\\
(T^{\mu\nu}_{\rm GRAV})_{\rm PART}+(T^{\mu\nu}_{\rm M})_{\rm PART}&=&0.
\label{M184}
\end{eqnarray}
All of the vacuum energy density infinities are taken care of by (\ref{M183}), and for astrophysics and cosmology we can then use the completely infinity-free (\ref{M184}). In this way, for studying white dwarfs or the cosmic microwave background,  in (\ref{M184}) we can now use $H=\sum a^{\dagger}(\bar{k})a(\bar{k})\hbar \omega_k$ alone after all, as the zero-point contribution has already been taken care of by gravity itself and does not appear in (\ref{M184}) at all. Moreover, when we do excite positive energy modes out of the vacuum we will generate an additional contribution to the cosmological constant, and it is this term that is measured in cosmology. Cosmology thus only sees the change in the vacuum energy density due to adding in positive energy modes and does not see the full negative energy mode vacuum energy density itself, i.e. in (\ref{M184}) one is sensitive not to  $\langle \Omega_M|T^{\mu\nu}_{\rm M}|\Omega_M\rangle$, and not even to  $\langle \Omega_M|bT^{\mu\nu}_{\rm M}b^{\dagger}|\Omega_M\rangle$ where $b^{\dagger}$ creates a positive energy mode out of $|\Omega_M\rangle$, but only to their difference  $\langle \Omega_M|bT^{\mu\nu}_{\rm M}b^{\dagger}|\Omega_M\rangle -\langle \Omega_M|T^{\mu\nu}_{\rm M}|\Omega_M\rangle$. Because of this one is able to find \cite{Mannheim2006} non-fine-tuned fits to the accelerating universe Hubble plot data  that are as good as the standard fine-tuned dark matter dark energy paradigm \cite{footnote7}.

Gravity's response to the change in vacuum energy density is mode by mode, i.e. gravity mode by matter mode. Thus in  applications of conformal gravity to cosmology the relevant  symmetry breaking field vacuum expectation value is not $\langle \Omega_M|S|\Omega_M\rangle$ but the altogether smaller $\langle \Omega_M|bSb^{\dagger}|\Omega_M\rangle- \langle \Omega_M|S|\Omega_M\rangle$ \cite{Mannheim2015,Mannheim2017}. In contrast, if one uses (\ref{M168}) as in Einstein gravity, disconcertingly gravity then sees an entire sum over matter modes and sees the full and large $\langle \Omega_M|S|\Omega_M\rangle$. To summarize, if one wants to take care of the cosmological constant problem, one has to take care of the zero-point problem as well, and when one has a renormalizable theory of gravity, via an interplay with gravity itself one is then able to bypass an equation such as (\ref{M168}) and take care of both of the two problems at one and the same time.


\begin{thebibliography}{99}
\bibitem{footnote1} If one starts with a path integral over both metric paths and matter field paths, formally (\ref{M167}) would correspond to a path integration over the matter fields and a stationary variation on the metric.  However, when one performs the path integration over all the other metric paths one finds that the path integral does not actually exist, with there being infinities not just in the vacuum zero-point sector but in scattering amplitudes as well.
\bibitem{footnote2} An initially attractive feature of a theory such as supersymmetry is that it partners fermions with bosons, and if the supersymmetry is exact zero-point fermion and zero-point boson contributions will identically cancel each other. However, this cancellation is lost once the supersymmetry is broken, and the lack to date of the discovery of any superparticles with masses below a few hundred or so GeV leads to a contribution to the vacuum energy density that is at least 60 or so orders of magnitude larger than (\ref{M168}) is able to handle.
\bibitem{footnote3} As long as one does not couple to gravity one is free to normal order. Thus for flat space electrodynamics as long as one can get to an analog of (\ref{M167}), which one can, one can proceed directly to an analog of (\ref{M168}). Since the electrodynamic analogs of (\ref{M167}) and (\ref{M168}) do follow from QED we see that there is no independent theory of CED. Rather, CED is output to QED.
\bibitem{footnote7} Unlike the standard model where acceleration is fine tuned to only occur at late times, in the conformal theory the Hubble plot is accelerating at all redshifts, a testable prediction.
\end{thebibliography}
\end{document}